\documentclass[twocolumn,showpacs,preprintnumbers,amsmath,amssymb]{revtex4}
\usepackage{graphicx}

\newcommand{\beq}{\begin{equation}}
\newcommand{\bey}{\begin{eqnarray}}
\newcommand{\eeq}{\end{equation}}
\newcommand{\eey}{\end{eqnarray}}
\def\grad{\nabla}

\def\kpc{\, {\rm kpc} }
\def\msun{M_\odot}

\def\lsim{\mathrel{\raise.3ex\hbox{$<$\kern-.75em\lower1ex\hbox{$\sim$}}}}
\def\gsim{\mathrel{\raise.3ex\hbox{$  $\kern-.75em\lower1ex\hbox{$\sim$}}}}

\def\kms{\, {\rm km \, s}^{-1} }

\def\r{{\mathbf r}}

\def\R{{\mathbf R}}

\begin{document}

\title{Timing and Lensing of the Colliding Bullet Clusters: barely enough time 
and gravity to accelerate the bullet}
\author{HongSheng Zhao}
\affiliation{University of St Andrews, Scottish University Physics Alliances, KY16 9SS, UK} \email{
hz4@st-andrews.ac.uk}
\begin{abstract}
We present semi-analytical constraint on the amount of dark matter in the 
merging bullet galaxy cluster using the classical Local Group timing arguments. 
We consider particle orbits in potential models which fit the lensing data.  
{\it Marginally consistent} CDM models in Newtonian gravity are found with a total mass 
$M_{CDM} = 1 \times 10^{15}\msun$ of Cold DM: the bullet subhalo can move
with $V_{DM}=3000\kms$, and the "bullet" X-ray gas can move with $V_{gas}=4200\kms$.  
These are nearly the maximum speeds that are accelerable by the gravity of 
two truncated CDM halos in a Hubble time even without the ram pressure.
Consistency breaks down if one adopts higher end of the error bars 
for the bullet gas speed ($5000-5400\kms$), and 
the bullet gas would not be bound by the sub-cluster
halo for the Hubble time.  Models with $V_{DM} \sim 4500\kms \sim V_{gas}$ 
would invoke unrealistic large amount $M_{CDM}=7\times 10^{15}\msun$ of CDM
for a cluster containing only $\sim 10^{14}\msun$ of gas.  
Our results are generalisable beyond General Relativity, e.g., a speed of $4500\kms$ 
is easily obtained in the relativistic MONDian lensing model of Angus et al. (2007).  
However, MONDian model with hot dark matter $M_{HDM} \le 0.6\times 10^{15}\msun$
and CDM model with a halo mass $\le 1\times 10^{15}\msun$ 
are barely consistent with lensing and velocity data.  
\end{abstract}

\pacs{98.10.+z, 98.62.Dm, 95.35.+d; submitted to Physical Review D, rapid publications}

\maketitle

\section{Potential from Timing}

Timing is a unique technique to establish the case for dark matter halos, 
first and most throughly explored in the context of the Local Group (Kahn \& Woljter 1959, 
Fich \& Tremaine 1991, Peebles 1989, Inga \& Saha 1998).  
In its simplest 
version the Local Group consists of the Milky Way and M31 as two isolated point masses,
which formed close to each other, moved apart due to the Hubble expansion, and 
slowed down and moved towards each other upto their present 
velocity $\sim 120\kms$ and separation (about 700 kpc)
due to their mutual gravity.  The age of the universe sets the upper limit on 
the period of this galaxy pair, hence the total mass of the pair through Kepler's 
3rd law assuming Newtonian gravity.

Timing also finds a timely application in 
the pair of merging galaxy clusters 1E0657-56 at redshift $z=0.3$, 
which is largely an extra-galactic grand analogy of the M31-MW system.  
The sub-cluster, called the "bullet", 
presently penetrates 400-700 kpc through the main cluster with an apparent speed of 
$\sim 4750^{+710}_{-550} \kms$
(Markevitch 2006).  The X-ray gas of the bullet (amounts to $2 \times 10^{13}\msun$)
collides with the X-ray gas of the main cluster (with the total gas up to $10^{14}\msun$)  
and forms a Mach-3 cone in front of the "bullet".  The two clusters have at least four 
different centers, which are offset by 400 kpc 
between the pair of X-ray gas centers and by 700 kpc between the pair of 
star-light centers, which coincides with the gravitational lensing centers and 
(dark matter) potential centers 
(Clowe et al. 2006).  The penetration speed is unusually high, hard for standard cosmology 
to explain statistically (Hayashi \& White 2006), and modified force law has been suggested
(Farrar \& Rosen 2006, Angus et al. 2007). 

The timing method applies in in MONDian gravity as well as Newtonian. Like lensing,
timing is merely a method about constraining potential distribution,  
and is only indirectly related to the matter distribution.
In this Letter we model the bullet clusters as a pair of mass concentrations 
formed at high redshift, and set constraint on their mutual force 
using the simple fact that their radial oscillation period must be 
close to the age of universe at $z=0.3$.   
We check the consistency with the lensing signal of the cluster
and give interpretations in terms of standard CDM and MOND.

First we can understand the speed of the bullet cluster analytically in simplified scenarios.  
Approximate the two clusters as points of fixed masses $M_1$ and $M_2$ on a head-on orbit, we can  
apply the usual MW-M31 timing argument.  The total mass $M_0=M_1+M_2$ is constant.  
The radial orbital period is computed from
\bey
T &=& 2 \int_0^{r_{max}} {dr \over V(r)}, \\  
  &=& 2 \pi \sqrt{r_{max}^3 \over G M_0}, \qquad \mbox{\rm Newtonian $p=2$} \\
 &=& {\sqrt{2\pi} r_{max} \over V_M},\qquad \mbox{\rm deep-MONDian, $p=1$} \\
 &\propto& K^{-n/2} r_{max}^{1+p \over 2}, \qquad \mbox{\rm for a $K/r^{p}$ gravity,} 
\eey
where 
$r_{max}$ is the apocenter and is related to the present relative 
velocity $V(r)$ at separation $r=700\kpc$ by energy conservation
\bey
{V(r)^2 \over 2} &=& -{GM_0 \over r_{max}} +{GM_0 \over r} \qquad \mbox{\rm Newtonian} \\
              &=& V_M^2 (\ln r_{max} - \ln r) \qquad {\rm deep-MONDian} \\
              &\propto & \left(r^{1-p} - r_{max}^{1-p}\right)K/(1-p) 
\qquad \mbox{\rm for a $K/r^{p}$ gravity,}
\eey
where $V_M = \sqrt{\xi} (G M_0 a_0)^{1/4}$ is the MOND circular velocity of 
two point masses, 
$a_0$ equals one Angstrom per square second and 
is the MOND acceleration scale, and the dimensionless
$\xi \equiv { 2 M_0^2 \over 3 M_1M_2 } 
\left( 1- \left({M_1\over M_0}\right)^{3/2}  - \left({M_2\over M_0}\right)^{3/2} \right) 
\sim 0.81 \sim 1$ 
(cf. Milgrom 1994, Zhao 2007, in preparation) for a typical mass ratio.

The predictions for simple Newtonian Keplerian gravity are given in Fig.~1; 
the more subtle case for a MONDian cluster is discussed in the final section.  
Setting the orbital period $T=10$Gyrs, the age of the universe at the cluster redshift, 
yields presently 
 $V \sim 3200\kms$ in Newtonian for a normal combined mass 
of $M_1+M_2=(0.7-1)\times 10^{15}\msun$ for the clusters, which is about 7-10 times 
their baryonic gas content ($\sim 10^{14}\msun$) 
for Newtonian universe of $\Omega=0.3$ cold dark matter.  
In agreement with Farrar \& Rosen and Hayashi \& White, the simple timing argument 
suggests that dark halo velocities of $4750\kms$, as high as the "bullet" X-ray gas, 
would require halos with unrealistically 
larger masses of dark matter, $\sim 10^{16}\msun$, an order of magnitude more than
what a universal baryon-dark ratio implies.  As a sanity check, 
assuming a conventional $3 \times 10^{12}\msun$ Local Group dark matter mass  
Fig.1 predicts the relative velocity of 
$\sim 100$km/s for the M31-MW system at separation 700 kpc after 14 Gyrs,
consistent with observation (Binney \& Tremaine 1987).

\begin{figure}
\includegraphics[angle=0,width=7cm]{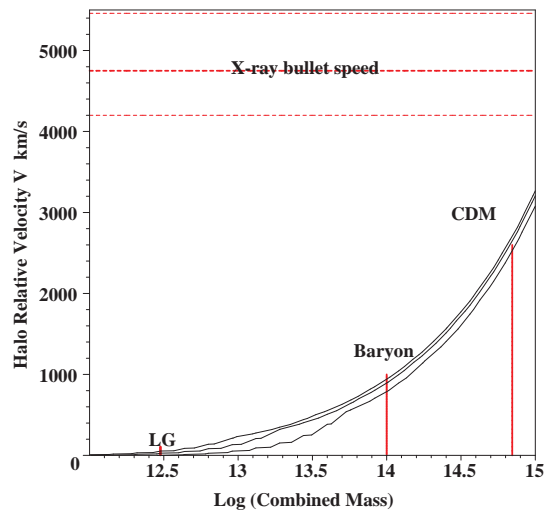}
\caption{Analytical timing-predicted dynamical mass vs. 
the relative speed of two objects separated by 700 kpc after $10 \pm 4$ Gyrs
(three lines in increasing order for increasing time) 
assuming Keplerian potential of point masses.  Three vertical lines indicate typical 
Local Group Halo mass, Baryonic mass in galaxy clusters, and most massive CDM halo masses. 
Three horizontal lines indicate the error bar of the speed of the X-ray "bullet" gas. 
}
\end{figure}
These analytical arguments, while straightforward, are not precise given its simplifying assumptions.
For one, clusters do not form immediately at redshift infinity, 
and the cluster mass and size might grow with time graduallly.  More important
is that point mass Newtonian halo models are far from fitting the weak lensing data of the 1E0657-56. 
A shallower Newtonian potential makes it even more difficult to accelerate the bullet.
On the other hand, Angus, Shan, Zhao, Famaey (2007) show that there are MOND-inspired
potentials that fits lensing.  As commented in their conclusion,  
the same potential is deep enough that a $V=4750\kms$ "bullet" is bound in an orbit 
of apocenter $r_{max}$ of a few Mpc, so the two clusters could be accelerated 
by mutual gravity from a zero velocity apocenter to 4750 km/s within the clusters' lifetime.  This line of thought was 
further explored by the more systematic numerical study of Angus \& McGaugh (2007).

Our paper is a spin-off of these works and the works of Hayashi \& White and 
Farrar \& Rosen.  We emphasize the unification of the semi-analytical timing perspective
and the lensing perspective, and aim to 
derive robust constraints to the potential, without being limited to a specific 
gravity theory or dark matter candidate.

Towards the completion of this work, we are made aware by the preprint 
of Springel \& Farrar (2007) that the unobserved bullet DM halo
could be moving slower than its observed stripped X-ray gas.
These authors, as well as the preprint of Milosovic et al. (2007), 
emphasized the effect of hydrodynamical pressure, which we will not 
be able to model realisticly here.  But to address the velocity differences, 
instead we treat the X-ray gas as a "bullestic particle".  
We argue that {\it our hypothetical ballistic particle must move 
slow enough to be bound to vicinity of the subhalo before the collision, 
but moves somewhat faster than $4700_{-550}^{+700}\kms$ now, 
since it does not experience ram pressure of the gas}.  
This model follows the spirit of classical 
timing models of the separation of the Large and Small Magellanic Clouds 
and the Magellanic Stream (Lin \& Lynden-Bell 1982). 

\section{3D Potential from Lensing}
 
The weak lensing shear map of Clowe et al. (2006) has been fitted by 
Angus et al. (2007) using a four-component analytical potential each being spherical 
but on different centres.  For our purpose we redistribute the minor components
and simplify the potential into two 
components centred on the moving centroid of galaxy light of the main cluster with the 
present spatial coordinates 
$r_1(t) =(-564, -176, 0)\kpc$ and subcluster galaxy centroid $r_2(t) = (145, 0, 0)\kpc$; 
the coordinate origin is set at the present brightest point of the "bullet" X-ray gas;
presently the cluster is at $z=0.3$ or cosmic time $t=10$Gyrs.  
We also apply a Keplerian truncation to the potential beyond the truncation radius $r_t$.  
So the following 3D potential is adopted for the cluster 1E0657-56 at time $t$,
\bey
\Phi(X,Y,Z,t)
            &=& (1800\kms)^2 \phi\left(|\r-\r_1|\right)\\\nonumber
            &+& (1270\kms)^2 \phi\left(|\r-\r_2|\right), \\
\phi(|\r-\r_i(t)|) &=& \ln \sqrt{1+\left({|r-r_i(t)|\over 180\kpc}\right)^2} + cst, ~r < r_t\\
        &=& - {\tilde{r}_t  \over |\r-r_i(t)|}, \qquad r \ge r_t(t)=C \times t,
\eey
where $\tilde{r}_t \equiv {r_t^3 \over r_t^2+ 180^2}$ is to ensure a continuous and 
smooth transition of the potential across the truncation radius $r_t$.  The truncation 
$r_t$ evolves with time, since a pre-cluster region collapses gradually after the big bang, 
and its boundary and total mass grows with time till it reaches the size of a cluster. 
In the interests of simplicity rather than rigour, 
we use a linear model $r_t = C \times t$, where $C$ is a constant of the unit kpc/Gyr. 

To check that the simplified potential 
is still consistent with weak lensing data, we recompute the 3D weak lensing convergence 
(Taylor et al. 2004) for sources at distance $D(0,z_s)$ at the redshift $z_s$, 
\beq
\kappa(X,Y,z_s)= \sum_{i=X,Y}
  {\partial_i \over 2} \left[\int_{0}^{D(0,z_s)} {2D(z,z_s)\over c^2} (\partial_i \Phi) dZ\right] 
\eeq
where the integrations in square backets are the deflection angles for a source at $z_s$, 
and the usrual lensing effective distance is related to the comoving distances by 
$D(z,z_s)=(1+z)^{-1} \tilde{D}(z) \left[ 1-{\tilde{D}(z) \over \tilde{D}(z_s)} \right] = 587$ Mpc 
is for the bullet cluster $z=0.3$ lensing sources at $z_s=1$; the distance increases by a factor 
1.3 to 1.6 for source redshifts of 3 to infinity.  Fig.2 shows the 
predicted $\kappa$ along the line joining the two dark centers; the result is insensitive
to the cluster truncation radius as long as $r_t \ge 1000$kpc presently.   
The lensing model predicts a signal in between that of the weak lensing data of 
Clowe et al., and strong lensing data of Bradac et al.  It is known that these 
two data sets are somewhat discrepant to each other.  So the fit here is reasonable.
The method is deprojection is essentially similar to the decomoposition method of Bradac et al.
whose explicit assumption of Einsteinian gravity is however unnecessary. 
\begin{figure}
\includegraphics[angle=0,width=7cm]{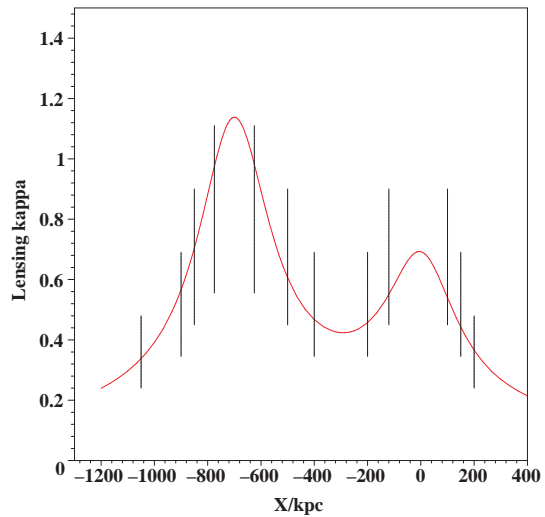}
\caption{Predicted bullet cluster convergence (rescaled for sources at infinity) 
along the line $Y=0.3X+cst$ connecting our two potential centroids.
The model predicts a lensing signal in between that of
observed weak lensing data from sources at $z_s=1$
(Clowe et al, lower end of error bars)
and the united weak lensing and strong lensing ($z_s=3$) data (Bradc et al. 
upper part of error bars); the mismatch of these two datasets are presently unresolved.}
\end{figure}

The important thing here is that as far as 
deprojecting the above potential is concerned, no assumption is needed on the gravity theory
as long as light rays follow geodesics, a feature built in most alternative gravity theory.
Similarly orbits of massive particles are also (different) geodesics in these theories.
The meaning of potential in such theories is that the potential
(scaled by a factor $2/c^2$) represents metric perturbations to 
the flat space-time, especially to the 
$g_{00}(cdt)^2=-(1+{2\Phi \over c^2})(cdt)^2$ term, so the Christoffel 
$\Gamma^i_{00} \sim {\partial \over \partial X^i} \Phi$,  it can be shown that
the geodesic equations have the same form as Einsteinian in the weak-field limit:
${d^2 \over dt^2}\R \approx - (1+{v^2 \over c^2})\grad_{\R} \Phi$, where 
$\R$ is the pair of spatial coordinates perpendicular to the instantaneous velocity $v$; 
the pathes of light rays are deflected twice 
as much by the metric perturbation $2\Phi/c^2$ as those of low-speed particles.

\section{Orbits of the colliding clusters}

We now use this potential to predict the relative speed of the two clusters.
This is possible using the classical timing argument, in the style of Kahn \& Wolter (159), 
Fich \& Tremaine (1991) and Voltonen et al. (1998);
we postpone most rigourous least action models
(Peebles 1989, Schmoldt \& Saha 1998) for later investigations since 
these require modeling a cosmological constant and other mass concentrations along 
the orbital path of the bullet clusters, which have technical issues in non-Newtonian gravity.  
We trace the orbits of the two 
centroids of the potentials according to the equation of motion 
${d^2\r_i \over dt^2} = -\grad \Phi(\r_i)$.  We assign different relative velocities 
presently (at $z=0.3$), and integrate backward in time and require the two centroids of 
the potential be close together at a time 10 Gyrs ago.  
The motions are primarily in the sky plane,
but we allow for 600 km/s relative velocity component in the line of sight.  
Clearly at earlier times when t is small, the two centroids are well-separated
compared to their sizes, so they move in the growing Keplerian potential of each other.
At latter times the centroids came close and move in the cored isothermal potential.

We shall consider models with a normal truncation 
$r_t = C \times t=1000\kpc$ at time $t=10$ Gyrs.  We also consider 
models with a very large truncation $C\times t=10000\kpc$.   
In the language of CDM, the truncation means the virial radius of the halo.
The present instantaneous escape speed of the model can be computed by 
$V_{esc}=\sqrt{-2\Phi(X,Y,Z,t)}$.  We find  
$V_{esc} \sim 4200-4500\kms$ in the central region of the shallower potential model 
with a present truncation $1000\kpc$.  The escape speed increases to 
$V_{esc} \sim 5700\kms$ for models with a present truncation $10000\kpc$. 

Fig.~3 shows the predicted orbits for different present relative velocities
$V_{DM}=|{d\r_2 \over dt} - {d\r_1 \over dt}|$.  
Among models with a normal truncation, we find $V_{DM} \sim 2950\kms$; a model 
with relative velocity $V_{DM} < 2800\kms$ would predict an unphysical orbital crossing at high redshift, 
while models with $V_{DM} > 3000 \kms$ would predict that the two potential centroids were never 
close at high redshift.  

Larger halo velocities are only possible in models with very large truncation.  
If the relative velocity is $4200\kms<V_{DM}<4750\kms$ between two cluster gravity centroids, 
then the truncation must be as big as $10$Mpc at $z=0.3$.

We also track the orbit of the bullet X-ray gas centroid
as a tracer particle in the above bi-centric potential.  
We look for orbits where the bullet X-ray gas will always be
bound to one member of the binary system since the ram pressure in a hydrodynamical 
collision is unlikely to be so efficient to eject the X-ray gas out of potential 
wells of both the main and sub-clusters.  This means that the bullet speed must not 
exceed greatly the present instantaneous escape speed of the model, which is 
$\sim 4200-4500\kms$ in the central region of the shallow potential of a model 
with a present truncation $1000\kpc$.  The escape speed increases to $\sim 5700\kms$ for 
models with a present truncation $10000\kpc$. 
The model with normal truncation is {\it marginally consistent} with the observed 
gas speed $V_{gas} \sim 4750_{-550}^{+710}\kms$.
The problem would become more severe if the potential were made shallower by an even 
smaller truncation.   The gas speed is less an issue in models with larger truncation.

In short the present velocity and lensing data are easier 
explained with potential models of very large truncation.  
Models with normal truncation have smaller gravitational power, can 
only accelerate the subhalo to $3000\kms$ in 10 Gyrs.  Models with normal CDM truncation 
can only accelerate the bullet X-ray gas cloud to $\sim 4200-4400\kms$, the escape speed,
{\it marginally consistent} with observations.

Above simulation results are sensitive to the present cluster separation, but
insensitive to the present direction of the velocity vector.  Unmodeled effects such as 
dynamical friction associated with a live halo will reduce the predicted $V_{DM}$
for the same potential, but the effect is mild since the actual 
collision is brief $\sim 0.1-0.3$Gyrs and the factor 
$\exp(-M^2/2)$ in Chandrasekhar's formulae sharply reduces dynamical friction
for a supersonic body, where $M \sim 2-3$ is the Mach number for the bullet.
\begin{figure}
\includegraphics[angle=0,width=8.5cm]{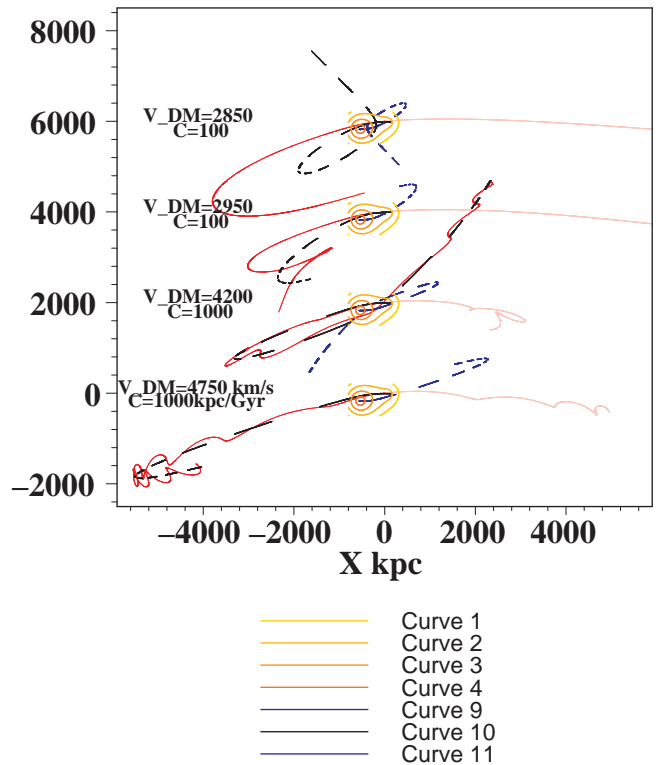} 
\caption{The orbit of the bullet subcluster X-ray gas (red, with present $V_{gas}=5400\kms$
for the 10 Gyrs in the past, and pink: for the future 4 Gyrs), and the orbits of the colliding main cluster halo 
(blue dashes) and subhalo (black dashes) in the potential (eqs. 8-10) 
determined by lensing data; dashes indicate length traveled in 0.5 Gyrs steps. 
No explicit assumption of gravity is needed for these calculations.
Orbits with different present halo relative velocity $V_{DM}$ 
and halo growth rate $C$ are shown after a vertical shift for clarity.
Timing requires the present cluster relative velocity in between $2800\kms<V_{DM}<3000\kms$
for potentials of normal truncation (lowest panels where the cluster truncation grows from zero to 
$C \times 10$Gyr $=1000\kpc$), and $4200\kms<V_{DM}<4750\kms$ for potentials with large truncation
(two upper panels where the cluster truncation grows from zero to 
$C \times 10$ Gyr $=10000\kpc$).}
\end{figure}

\section{Newtonian and MONDian meanings of the potential model}

Assuming Newtonian gravity the models with
normal truncation $r_t=1$Mpc at $t=10$Gyrs correspond to cluster (dark) masses of 
$M_1=0.745 \times 10^{15}\msun$ and 
$M_2=0.345\times 10^{15}\msun$; the larger truncation $r_t=10$Mpc corresponds to 
$M_1=7.45 \times 10^{15}\msun$ and 
$M_2=3.45\times 10^{15}\msun$ in Newtonian.  All these models fit lensing.

Interpreted in the MONDian gravity, the truncation is due to external field effect and  
cosmic background so to make the MOND potential finite hence escapable 
(Famaey, Bruneton, Zhao 2007).  Beyond the truncation radius, MOND potential becomes 
nearly Keplerian.  The MONDian models, insensitive to truncation, would have masses
only $M_1=0.66\times 10^{15}\msun$ and $M_2=0.16\times 10^{15}\msun$.  These masses are
still higher than their baryonic content $\sim 10^{14}\msun$, 
implying the need for, e.g., massive neutrinos; the neutrino 
density is too low in galaxies to affect normal MONDian fits to galaxy rotation curves,
but is high enough to bend light and orbits significantly on 1Mpc scale.  
The neutrino-to-baryon ration, approximately 7:1 in the bullet cluster, 
would be a reasonable assumption for a MONDian universe with $\Omega_b \sim 0.04$ plus
2eV neutrinos hot dark matter $\Omega_{HDM} \sim 0.25 \sim 7 \times \Omega_b$ 
(Sanders 2003, Pointecoute \& Silk 2005, Skordis et al. 2006, Angus et al. 2007).
The amount of hot dark matter inferred 
here is the same as Angus et al. (2007) since their potential parameters are fixed 
by the same lensing data.

\section{Conclusion}

In short a consistent set of simple lensing and dynamical model of 
the bullet cluster is found.  The present relative speeds between galaxies 
of the two clusters is predicted to be $V_{DM} \sim 2900\kms$ in CDM 
and $V_{DM} \sim 4500\kms$ in $\mu$HDM (MOND + Hot Dark Matter) 
if the two clusters were born close to each other 10 Gyrs ago; both 
models assume close to universal gas-DM ratio in clusters, i.e., about 
$(0.6-1)\times 10^{15}\msun$ Hot or Cold DM.   
Modeling the bullet X-ray gas as ballistic particle, we find 
the gas particle with speed of $V_{gas}=4200$km/s (at the lower end of observed 
speed) is bound to the potential of the subcluster
for most part of the Hubble time for both above models, insensitive to the preference
of the law of gravity.  But if future relative proper motion measurements 
of the subcluster galaxy speed is as high as $V_{DM}=4500$km/s, or the gas 
speed is as high as $V_{gas} \sim 5400\kms$, then  
Newtonian models would need to invoke unlikely $7 \times 10^{15}\msun$ DM halos
around $10^{14}\msun$ gas.

\end{document}